# Light pollution is skyrocketing

A new citizen science based study shows a worrying increase of light pollution

*Fabio Falchi[1,2] and Salvador Bará[3]*

**Artificial light at night is a pollutant that is rising fast, as demonstrated by Kyba et al. (1) work by analyzing ten of thousands observations by citizen scientists in the last 12 years. The study found that the dimmest stars are vanishing, progressively hidden by a 10% yearly increase of the sky background due to artificial lights. This increase is difficult to be detected by the global coverage satellites now in operation, due to detector's blindness to the blue peak of white LEDs that are progressively replacing older technology lamps. This shows the need for a satellite with nighttime multi band capability in the visible light to study and control future evolution. More importantly, a call for a strong reverse in the light pollution rising trend is extremely urgent to avoid all the cultural, scientific, energetic, ecological and health negative effects of artificial nightlights.**

Kyba et al. studied the collection of observations of Globe at Night project (https://www.globeatnight.org). The more than 50,000 observations, done by citizens using the naked eyes to find the dimmer stars visible in a site, allow to retrieve the background brightness of the night sky, as the brighter the night sky is, the brighter a star should be to be seen. The analysis of the dataset for the last dozen years shows that the sky brightness due to artificial light is increasing exponentially in the World with an alarming average of 10% each year, so doubling in less than 8 years. This increase is much higher than the estimates of the evolution of artificial light emissions, about 2% yearly, based on DMSP and Suomi NPP satellites radiance measurements (2,3). Part of this discrepancy could be explained by the impossibility of these satellites to detect the blue light, emitted in great quantity by the LEDs light that started to be used outdoors about 10 years ago. These satellites are also not able to see well the light emitted mainly horizontally, such as that from the increasing number of ultra-bright LED billboards and lighted buildings' façades. These weak points of the now operating satellites should be urgently overcome with a next generation satellite dedicated to light pollution, to allow for multi band and multi-angle monitoring capabilities (4,5) that this global environmental problem (6,7) deserve.

The most important message that the scientific community should catch from Kyba et al. study is that light pollution is increasing, notwithstanding the rising awareness on the problem and the

1 Departamento de Física Aplicada, Universidade de Santiago de Compostela; Santiago de Compostela, Galicia, Spain. E-mail: falchi@istil.it
2 ISTIL – Light Pollution Science and Technology Institute, Thiene, Italy
3 Independent Scholar; Santiago de Compostela, Galicia, Spain

countermeasures purportedly put in operation to limit it. The awareness has to increase much, for artificial light at night to be perceived by people not as an always positive thing, but as a pollutant, as it really is. Looking at the International Space Station's images and videos of the artificially lighted Earth's night hemisphere, people and even the astronauts are only able to see the 'beauty' of the city lights, as they were lights of a Christmas' tree. They don't perceive that these are images of pollution. It is like admiring the beauty of the rainbow colors that gasoline produces in water and not recognize that it is an image of chemical pollution.

People, media and politicians are used to associate to artificial light thaumaturgical properties on road safety and personal security that it seems not to merit (8,9). So, year after year more and more light is installed to light up the night. This does not come for free. Negative consequences are rising and new ones discovered. Astronomers were the first to realize that there was a problem with night light, and they progressively moved their observatories farther and farther from cities. This strategy worked for some time, but now they cannot go farther yet, while lights are approaching. In fact, due also to the increasing efficiency of light technology, that allows to get more light for the same money, roads that once were not lighted, now flourish with lampposts. To make things worse for astronomers, now the attack is coming also from space, where corporations are sending thousands of satellites that for their own monetary advantage, will prevent all humankind to admire a pristine sky anywhere on Earth. The wilderness experience one can have in a National Park during the day, most probably will not be repeated at night, due to the far glows of cities and now also due to the contamination by satellites megaconstellations. The loss of the starry night is an unprecedented loss for all cultures.

Light pollution is also, and mainly, an environmental problem (10). Life on Earth evolved with sunlight during the day and starlight and the Moon, when present, during the night. If we introduce in ecosystems artificial light to levels that surpass, even by thousands of times and more, the level experienced in natural conditions, animal behavior will change consequently. Some species, e.g. some predators, may take advantage, albeit temporarily, by more light. Others will be in danger of local or global extinction, diminishing progressively our local and global biodiversity (11). Moreover, animals, humans included of course, when exposed to artificial light at night diminish or stop their production of melatonin (12), a fundamental hormone for our internal clock, with cascade negative consequences on health (e.g. 13). Not to be forgotten is the contribution to global warming of greenhouse gases produced by installing, operating, maintaining, dismantle and recycle all types public and private outdoor lighting. An estimated 200 billion kg of carbon dioxide is generated to produce the ~400 TWh of electric energy needed for outdoor lights (14).

Attempts of control of light pollution have been carried out in the last decades in several places, at local up to national level. These efforts were not able to stop the increase of artificial light produced, even if, in some cases limited the consequences of this increase by installing sources that pollute less, for example by aiming the light flux only below the horizon plane. This approach is not sufficient, as any new light, even if shielded, will add pollution to the night environment after being reflected off the surfaces intended to be lit. New approaches that treat light as all other pollutants by introducing total caps and red-lines based on the control of indicators are urgently needed to start reducing light pollution (15,16), analogously to the reduction of most air contaminants obtained in the past decades thanks to legislative interventions, like the Clean Air Act.

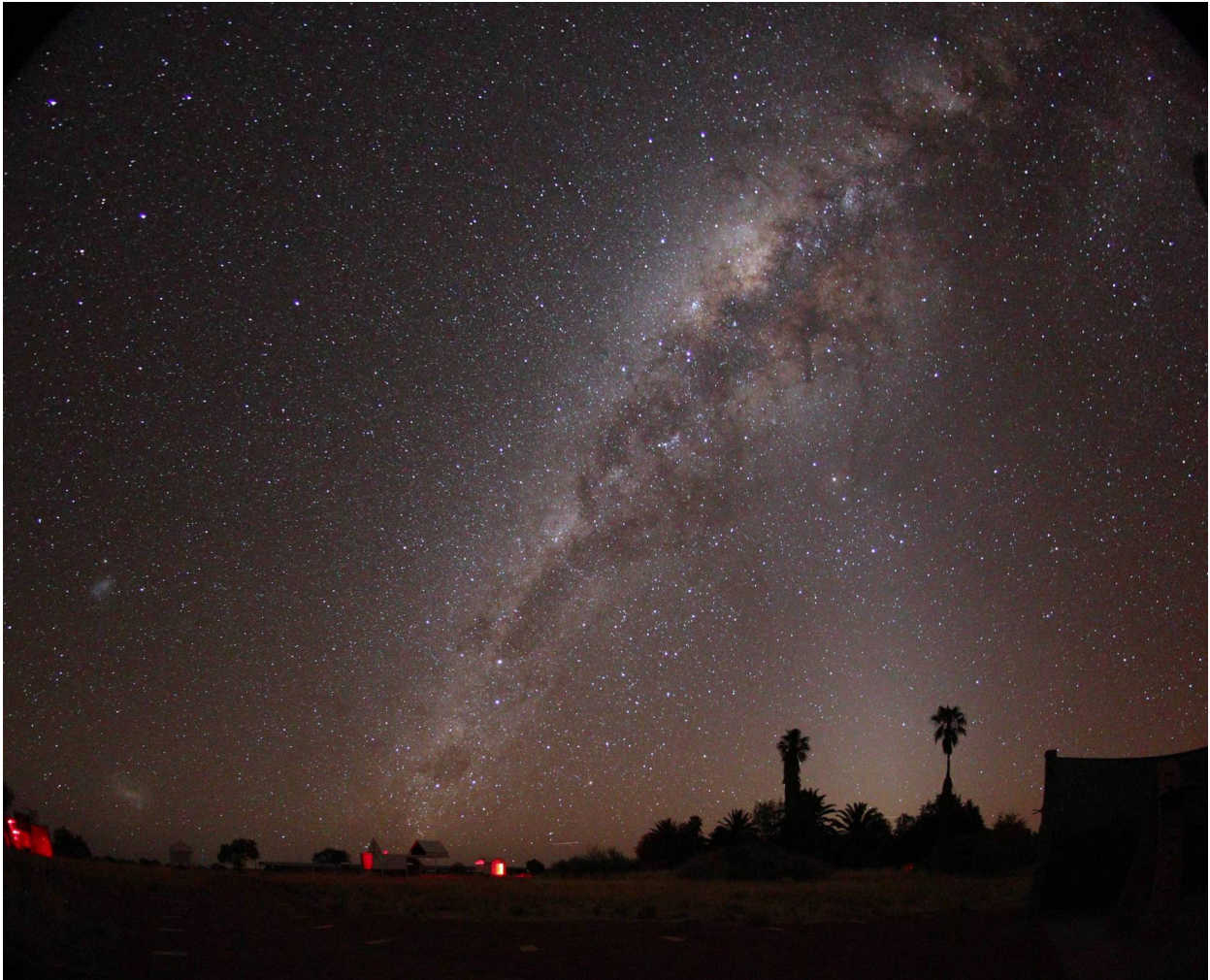

An uncontaminated night sky. The starry sky, as seen from Tivoli Astro Farm in Namibia, shows the greatest wonder of Nature with an eye-catching Milky Way, the Zodiacal light and the Magellanic clouds. This sky is now lost for the great majority of humankind (Photo by F.Falchi).